\documentclass[manuscript]{aastex}

\usepackage{epsfig,graphicx}
\usepackage{natbib}
\bibpunct{(}{)}{;}{a}{}{,}    

\begin{document}

\title{BIPOLAR MAGNETIC REGIONS ON THE SUN: \\ GLOBAL ANALYSIS OF THE
  \textit{SOHO/MDI} DATA SET}

\author{ J. O. Stenflo$^{1,2}$  and A.G. Kosovichev$^{3}$} 
\affil{$^1$Institute of Astronomy, ETH Zurich, CH-8093 Zurich, Switzerland }
\affil{$^2$Istituto Ricerche Solari Locarno, Via Patocchi, 6605 Locarno-Monti, Switzerland}
\affil{$^3$Stanford University, Stanford, CA 94305}

\email{stenflo@astro.phys.ethz.ch; AKosovichev@solar.stanford.edu} 

\begin{abstract}
The magnetic flux that is generated by dynamo processes inside the Sun
emerges in the form of bipolar magnetic regions. The properties of these directly
observable signatures of the dynamo can be extracted from  
full-disk solar magnetograms. The most homogeneous, high-quality synoptic data set 
of solar magnetograms has been obtained with the MDI instrument on the
SOHO spacecraft during 1995-2011. We have developed an IDL program
which has, 
when applied to the 73,838 magnetograms of the MDI data set, 
automatically identified 160,079  bipolar magnetic regions that span
a range of scale sizes across nearly four orders of magnitude. The
properties of each region have been extracted and statistically
analysed, in particular with respect to the polarity orientations of
the bipolar regions, including their tilt angle distributions and
their violations
of Hale's polarity law. The latitude variation of the average tilt
angles (with respect to the E-W direction), which is known as Joy's
law, is found to closely follow the relation $32.1^\circ \times 
\sin$(latitude). There is no indication of a dependence on region
size that one may expect if the tilts were produced by
the Coriolis force during the buoyant rise of flux loops from the
tachocline region. A few percent of all regions have orientations that violate
Hale's polarity law. We show explicit examples, from different phases
of the solar cycle, where well defined medium-size
bipolar regions with opposite polarity orientations occur side by side
in the same latitude zone in the 
same magnetogram. Such oppositely oriented large bipolar regions
cannot be part of the same toroidal flux system, but different flux
systems must coexist at any given time in the same 
latitude zones. These examples are incompatible with
the paradigm of coherent, subsurface
toroidal flux ropes as the source of sunspots, and instead show that
fluctuations must play a major role 
at all scales for the turbulent dynamo. To confirm the profound role
of fluctuations at large scales we show
explicit examples where large bipolar regions differ from the average
Joy's law orientation by an amount between $90^\circ$ and
$100^\circ$. We see no observational 
support for a separation of scales or a division between a global and
a local dynamo, since also the smallest scales in our sample retain a
non-random component that significantly contributes to the accumulated
emergence of a north-south dipole moment that will lead to the replacement of
the old global poloidal field with a new one that has the opposite orientation. 
\end{abstract}

\maketitle


\section{Introduction}\label{sec:intro}
The origin of solar activity with its 11 yr cycle is generally
understood in terms of an oscillatory dynamo inside the Sun
\citep[cf.][]{stenflo-brandsub05,stenflo-brandenburg05}. While a 
toroidal magnetic field is generated from a poloidal field through 
differential rotation, the essence of the global dynamo is the
regeneration of the poloidal field from the toroidal one through
interaction with turbulent convection in the rotating
Sun. The systematic twisting due to the Coriolis force generates an N-S
dipole moment from the originally E-W oriented toroidal field. This dipole moment
spreads by turbulent diffusion to reverse the old global poloidal 
field and replace it with a new one of opposite sign. 

This general scenario provides a framework for the explanation of
various observed properties of 
the cycle: (1) Hale's polarity law \citep{stenflo-haleetal19}, according to
which sunspots appear in pairs approximately aligned with the E-W
direction, but with opposite polarity orientations in the two
hemispheres, and with reversal of the orientations between two
successive 11-yr cycles. (2) The butterfly diagram, according to which
the latitude zones, in which the sunspots appear, migrate from high
($\sim 40^\circ$) to low (near the equator) latitudes as the cycle
progresses. (3) Joy's law \citep[also reported in the seminal paper
by][]{stenflo-haleetal19}, according to which the 
orientations of the magnetic polarities of the sunspot groups deviate
from the E-W direction in a systematic way, such that the preceding
(W) part of the region is on average closer to the equator than the
following (E) part. 

While much of the empirical studies of the solar cycle have been in
terms of the properties and distributions of sunspots, we know that sunspots only
represent proxies for the underlying fundamental agent, the magnetic 
field. The solar dynamo operates in the Sun's interior, which is
not accessible to direct observations (except indirectly through
helioseismology). At the surface of the Sun the main directly observable signatures of the
dynamo are the bipolar magnetic regions, which represent magnetic flux
that has emerged from the interior, and the large-scale flux patterns that
are shaped by the accumulated patterns of flux emergence combined with
the action of flux dispersion and other transport processes (like
meridional circulation). 

Bipolar magnetic regions occur at all scales with a size spectrum
that follows a power law
\citep{stenflo-harveyphd93,stenflo-harveyzwaan93,stenflo-schrharv94,stenflo-parnelletal09}, from the largest regions that 
harbor major sunspots often accompanied by violent flare activity, to
ephemeral active regions \citep{stenflo-harveymartin73,stenflo-harveyetal75,stenflo-martinharvey79} without sunspots, and
the still smaller intranetwork fields. The global contribution to the
overall flux emergence rate dramatically increases as we go down in
scale size \citep{stenflo-zirin87}, suggesting that the global flux balance is
dominated by the smallest scales, in contrast to the general visual
impression that one may get from magnetograms. 

Also the smaller
scales seem to statistically follow Hale's polarity law and the
butterfly diagram distribution, although with increasing statistical spread as we
go down in scale. It is not yet known at what scales the polarity
orientations become so randomized that the accumulated effect of
still smaller scales no longer contribute significantly to the
operation of the global dynamo. Magnetoconvection is expected to cause
magnetic structuring down to scales of order 10-100\,m, where the magnetic
Reynolds number becomes unity so that the magnetic field ceases
to be frozen-in and decouples from the turbulent plasma
\citep{stenflo-dewijn09}.  

The observational signature of the dynamo mechanism that most directly 
represents the regeneration of the poloidal field from the toroidal
one is the systematic tilt of the bipolar magnetic regions that is
statistically described by Joy's law. Combined with Hale's polarity
law this tilt describes how the emerging bipolar regions bring to the
surface an N-S bipolar moment that is the seed for the regeneration of
the new global poloidal field of reversed polarity, from which the
subsequent 11 yr cycle of solar activity is generated. The observed
properties of the tilt angles therefore give guidance to the dynamo
theories and constrain the ways in which the dynamo is allowed to
operate. 

The tilt angles of sunspot pairs have been studied in many papers
since the discovery by \citet{stenflo-haleetal19} of Joy's law, using
Kitt Peak magnetograms \citep{stenflo-wangsheeley89}, Mount Wilson magnetograms \citep{stenflo-howard91a}, 
and sunspot group data \citep{stenflo-howard91b}, confirming the overall
magnitude and crude latitude dependence originally found by
\citet{stenflo-haleetal19}. In addition \citet{stenflo-wangsheeley89}
found that 4\,\%\ of the bipolar magnetic regions had inverse polarity
orientations, and that the spread in the distribution of tilt angles
increased significantly as one goes to smaller regions. The
superb quality of the SOHO/MDI data set will allow us in the present
work to explore these
properties with much better precision and in much greater detail than
has been possible before. 

Recently \citet{stenflo-ks08} used the set of MDI magnetograms to
study the tilt changes in emerging bipolar magnetic regions. They
found that during the first few days after emergence the tilt angles relaxed towards the value
expected from Joy's law and not towards the E-W orientation, in 
agreement with the findings of \citet{stenflo-sivaramanetal07} from an
analysis of Kodaikanal white-light images. \citet{stenflo-ks08}
further found no dependence of the tilt behavior on the amount of flux
or size of the bipolar regions. Both these findings contradict the paradigm
that the tilt is caused by the Coriolis force acting on initially
untilted flux loops that rise from a toroidal source
region near the bottom of the convection zone and emerge at the
surface as tilted bipolar regions
\citep{stenflo-dsilva93,stenflo-fisheretal95}. Instead the tilt,
which is the source of the N-S dipole moment that leads to the
reversal and 
regeneration of the poloidal field, appears to have been 
established already in the dynamo region in the Sun's
interior. The tilt observed at the surface reflects this property
regardless of the size or amount of flux of the observed regions.


\section{Automatic identification of bipolar magnetic regions}

\subsection{Data set}
We have made use of the complete set of 96 minute cadence SOHO/MDI
full disk magnetograms \citep{stenflo-scherreretal95}, which covers
the 15 year period May 1996 -- 
April 2011. With a pixel size of $2\times 2$\,arcsec$^2$, the
effective spatial resolution of the magnetograms is $4\times
4$\,arcsec$^2$. The magnetograms represent maps of the line-of-sight component
of the magnetic flux density averaged over the spatial resolution
window. They have been derived from maps of the circular polarization
recorded with a narrow-band filter at different wavelength positions
within the Ni {\sc i} 6768\,\AA\ line \citep{stenflo-scherreretal95}. 

The complete data set comprises 73,838 full disk magnetograms. Some of the
magnetograms have defects (with pixel values NaN, which we replace by
zero). We have rejected from further analysis the 1572 magnetograms
that have more than 100 defect pixel values (out of a million for each
magnetogram). We have further rejected the 2272 magnetograms for which
the $P$ angle is neither (within one degree) $0^\circ$ (implying that heliographic N is
upwards, along the $y$ axis) nor $180^\circ$ (with heliographic S in
the upwards direction), to avoid having to deal with odd rotations of the
coordinate system. An additional 60 magnetograms have been rejected
because the value of the Julian date in the header is inconsistent
with other header information.

\subsection{Limitations, reliability, and incompleteness}
Our aim has been to develop a computer program that can be applied to
any of the 73,838 MDI magnetograms, to automatically identify the
bipolar magnetic regions and extract their properties. However, it
does not seem feasible to make a program that can automatically and
reliably  identify {\it 
  all} the bipolar regions. The algorithm that we have developed therefore
makes trade-offs between 
reliability and completeness. Our priority has been 
reliability, at the expense of completeness. The regions identified by
the program should be truly
individual bipolar magnetic regions, rather than clusters of several
bipolar regions during times of high solar activity, or chance
encounters between opposite polarities in less active areas. By
choosing the criteria for an identification in a conservative way to
minimize false identifications, the program will fail to identify many
truly bipolar regions. Although our set of bipolar regions will
represent a sufficiently reliable sample, it is a sample that is
incomplete. 

The automatic identification of the bipolar regions is done for each
magnetogram separately, without following the evolutionary history of
the magnetic patterns from one magnetogram to the next. Therefore the
same bipolar regions may be identified in different magnetograms but
formally treated as separate, and different 
regions may represent different evolutionary phases. Thus the
measured region properties may include evolutionary effects, which can 
add to the spread in the distribution functions of their 
properties. However, since only relatively compact and well defined
bipolar regions that have not yet significantly decayed get identified
by the program, processes like flux transport and rotational shearing
should not affect the results much. 

It is known from previous studies that the number of bipolar magnetic
regions increases with decreasing region size according to a power law
\citep{stenflo-harveyphd93,stenflo-harveyzwaan93,stenflo-schrharv94,stenflo-parnelletal09}. This implies that the majority of
regions are small, near the resolution limit of the MDI
magnetograms. However, as we need to set the identification criteria
to reliably extract the largest bipolar regions during times of maximum solar activity,
when the magnetograms are very crowded, the identification thresholds
need to be set so high that a large fraction of the smallest regions
will be missed. In spite of these compromises the program succeeds, as
we will see, to extract regions over the full range of scales that
spans several orders of magnitude in flux. This allows us to explore
how the properties of the bipolar regions depend on their sizes and 
fluxes. 

\begin{figure*}
\resizebox{\hsize}{!}{\includegraphics{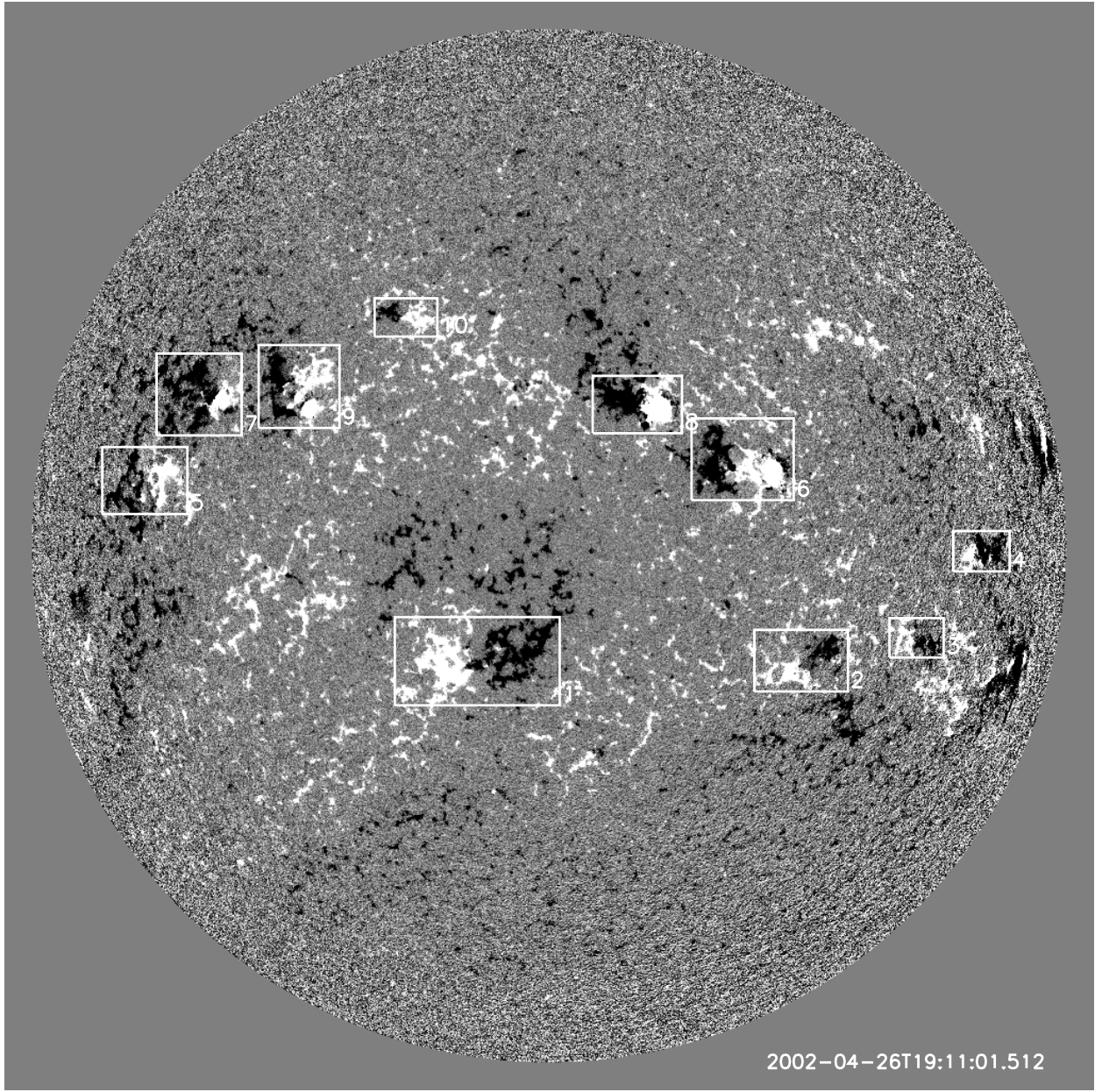}}
\caption{Vertical flux density map, calculated from the MDI line-of-sight
magnetogram recorded on April 26, 2002, as described in Appendix
A.1. The IDL program has identified 10 bipolar magnetic 
  regions and enclosed them in rectangular boxes. Only 8 of them are
  retained for the analysis, since the two boxes to the far left and
  the far right do not satisfy the criterion of limb zone
  avoidance. The grey scale cuts are set at +100\,G (white) and
  $-100$\,G (black). 
}\label{fig:fulldisk}
\end{figure*}

Figure \ref{fig:fulldisk} shows an example of one of the many 
magnetograms, with 10 identified bipolar magnetic regions enclosed by
their respective rectangular boxes. Since the program rejects bipolar
regions that are too close to the solar limb (see Appendix, 
Sect.~\ref{sec:steps}), only 8 of these regions
have been retained for further analysis. 

\subsection{Set of extracted parameters}\label{sec:extract}
A detailed description of the reduction steps with the set of
selection criteria that must be satisfied for the identification of
acceptable bipolar magnetic regions is given in
the Appendix. After the IDL program has
looped through all the 73,838 magnetogram files, it has written an IDL
save file with the extracted bipolar region information for each of
the analysed magnetograms. These save files are then merged into one
single master file that contains the extracted information for a 
total of of 245,733 identified bipolar regions, together with the
relevant housekeeping data, like time of the magnetogram, heliographic
longitude of the central meridian, heliographic latitude of disk
center, Carrington rotation number, etc. 

From this data set we remove the bipolar regions, for which the
derived centers of gravity do not lie within $r/r_\odot <0.8$ for {\it
both} polarities, in order to avoid errors that can occur for regions
that are located close to the solar limb. $r/r_\odot =0.8$ corresponds
to the cosine of the heliocentric angle $\mu=0.6$. This limb zone is
conservatively chosen to 
be this wide to minimize sources of error due to projection effects and 
noise. 85,654 regions, or 
34.9\,\%\ of the total, do not satisfy this criterion. This leaves us
with a rest of 160,079  bipolar regions, which satisfy all our
criteria and enter into our statistical analysis. 

For each of these bipolar magnetic regions the heliographic
coordinates for the centers of gravity and the total magnetic fluxes
of each of the two magnetic polarities are determined and saved, accounting for
the geometric foreshortening effect for pixels away from disk
center. Using spherical trigonometry, we connect the centers of
gravity of the two polarities with a great circle and calculate the
tilt angle of this great circle with respect to the parallel circle
that defines the E-W direction, as well as the polarity separation (in Mm along the great
circle). 

For the N hemisphere the tilt is
defined to increase in the clockwise direction, for the S hemisphere
in the counter-clockwise direction. The zero point of the tilt angle
represents a bipolar region aligned along the E-W direction with the
preceding (westward) polarity being of positive sign. The tilt angle
for regions with opposite orientation thus differs by 180$^\circ$.


\section{Tilt angles}
Our set of 160,079  bipolar regions provides us with a rich data base
with which we can test and explore in detail the behavior of Joy's
law. According to this law the orientations of the bipolar magnetic
regions are tilted with respect to the E-W direction, such that (on
average) the preceding (westward) polarity is closer to 
the heliographic equator than the following polarity. This systematic
tilt plays a fundamental role for any dynamo
theory of the solar activity cycle, since it implies the presence of N-S dipole
moments of the bipolar regions, which represent seeds for the reversal
and regeneration 
of the Sun's poloidal magnetic field. 

The Coriolis force acting on
buoyant toroidal flux 
elements rising from the Sun's interior is generally invoked as the
cause of the tilt with its opposite signs in the two hemispheres. With 
this process the
magnitude of the tilt is governed by the strength of the toroidal
field in the storage region inside the Sun and the rise time of the
flux loops to the
surface. As we will see, our present analysis does not support this
scenario. 

The tilt angles have a strong and well defined latitude
dependence. Our analysis however indicates that there is no
significant variation of the average tilt angles with phase of the
solar cycle \citep[as also found by][]{stenflo-wangsheeley89} or with
the amount of flux (cf. Sect.~\ref{sec:flux} below), 
although the statistical spread of tilt angles increases when we go to smaller
regions. Therefore we will start with a global analysis, where we
lump the data for the whole 15 yr MDI data set together, to explore
latitude and size variations with optimized statistics, while ignoring
the possibility of subtle temporal variations of Joy's law (for which
there seems to be no evidence).

\subsection{Latitude dependence}
To establish a well-defined reference
relation for Joy's law with optimized statistics, we use all bipolar regions for both
hemispheres and all phases of the solar cycle together. Since for
approximately half the regions the 
positive polarity is the preceding polarity, while for the other
half it is the following polarity, we get a bimodal angular
distribution, with two peaks separated by $180^\circ$. Since these
peaks are identical except for their $180^\circ$ separation, we bring
them on top of each other by subtracting $180^\circ$ from tilt angles
that fall within quadrants 3 and 4 ($90^\circ -270^\circ$, to make all
angles fall within quadrants 1 and 2. This gives us a single peak for
the angular distribution, which we can fit with a Gaussian to
determine its position and spread. 

The analysis has been made for a set of 9 latitude bins, spanning the range
$0^\circ -45^\circ$ with a bin width of $5^\circ$. The corresponding
latitude range for the S hemisphere has been mapped on top of these positive
latitude bins, since there is no evidence for any difference in the
behavior of Joy's law between the two hemispheres except for the sign
change of the tilt. We recall that the tilt angles that we have determined
for the N hemisphere are defined to be positive if the tilt is
clockwise, while those in the S hemisphere are positive if the tilt is
counter-clockwise. 

\begin{figure}
\resizebox{\hsize}{!}{\includegraphics{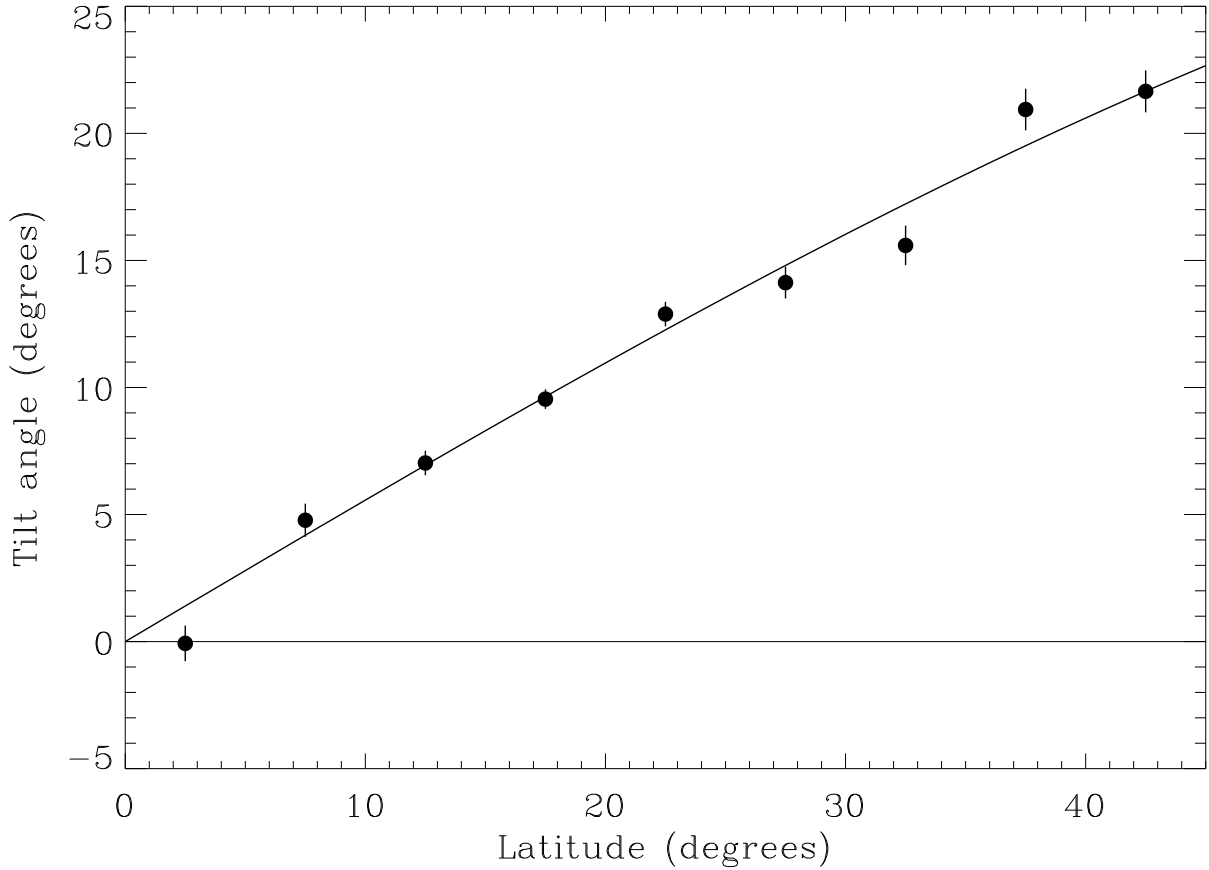}}
\caption{Average tilt angle as a function of latitude, for all bipolar
  regions with the N and S hemispheres combined. Tilt angles outside
  the range [$-90^\circ,+90^\circ$] have been shifted by $180^\circ$
  to be brought back to this interval, to allow all regions to be
  described by a single relation, irrespective of hemisphere or
  cycle. Positive tilt angle $\gamma$ means that the preceding polarity is
  equatorwards of the following polarity. The solid curve is 
  the fit function $\gamma = 32.1^\circ \,\sin b$, where $b$ is the
  heliographic latitude. 
}\label{fig:globaltilt}
\end{figure}

The result for the mean tilt angles in each latitude bin, as derived
from a Gaussian-type fit to the angular distributions, is shown in
Fig.~\ref{fig:globaltilt} as the solid circles with their respective
1-$\sigma$ error bars. Through this set of points we have fitted the
analytical function 
\begin{equation}
\gamma = \gamma_0 \,\sin b\,,\label{eq:globaltilt}
\end{equation}
where $b$ is the heliographic latitude, and the tilt amplitude
$\gamma_0$ is the single free fit parameter. This is the natural choice of fit
function if the origin of the systematic tilt  is related to the
Coriolis force, since this force varies 
with latitude as $\sin b$. The tilt amplitude is found to be 
\begin{equation}
\gamma_0 = 32.1^\circ\pm 0.7^\circ\,.\label{eq:tiltamp}
\end{equation}

This rather large tilt angle amplitude is in good general agreement
with the results of \citet{stenflo-wangsheeley89}, but it disagrees
with the much flatter latitude dependence found from Mount Wilson and
Kodaikanal data by \citet{stenflo-espuig11} and deduced from numerical simulations by
\citet{stenflo-schbau06}.

\subsection{Angular distribution}

\begin{figure}
\resizebox{\hsize}{!}{\includegraphics{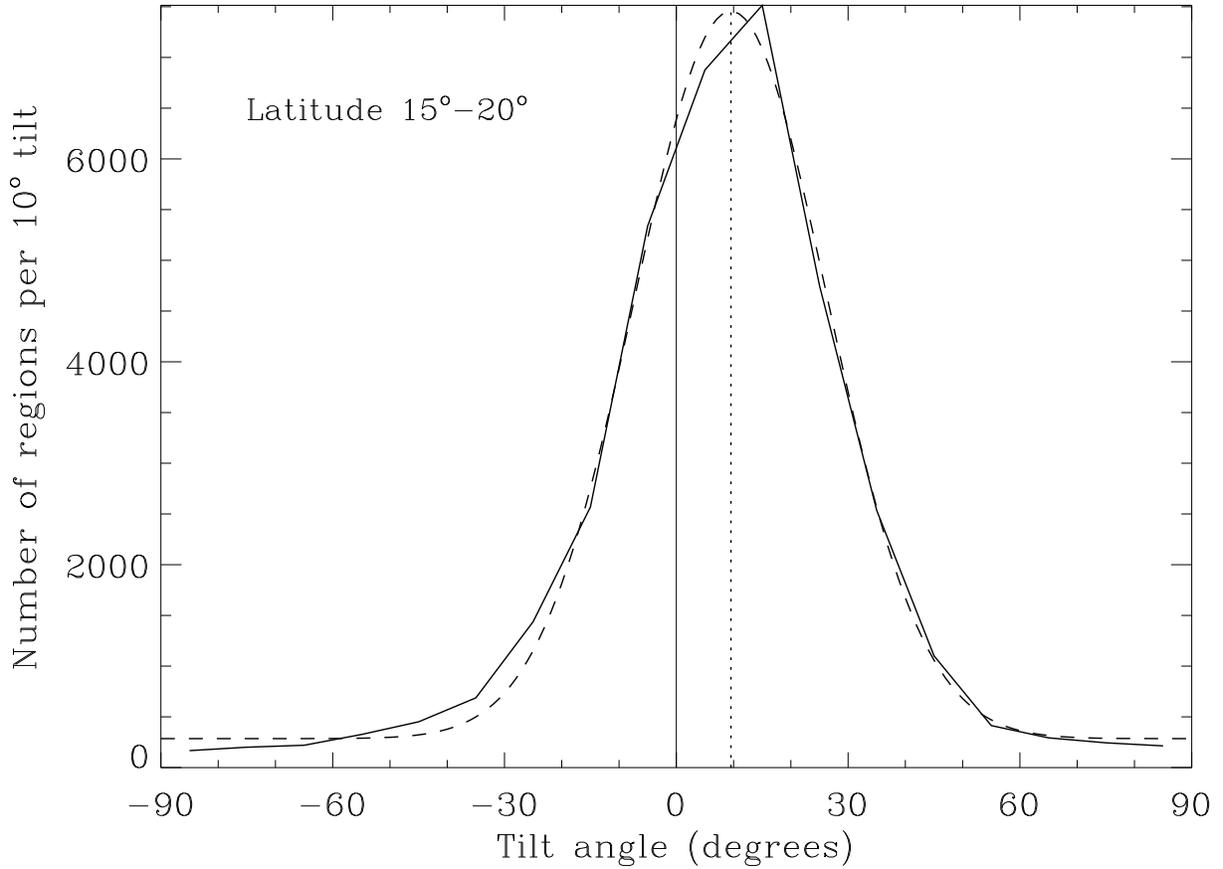}}
\caption{Histogram for the angular distribution of the tilt angles in
  the latitude zone $15^\circ$ - $20^\circ$ for both
  hemispheres combined, like in Fig.~\ref{fig:globaltilt}. The dashed
  curve represents a fit with a Gaussian-type function, centered around the
  angle marked by the dotted line. 
}\label{fig:angdist}
\end{figure}

Figure \ref{fig:angdist} illustrates what the angular distributions for
the tilt angles look like, here for the latitude bin $15^\circ$ -
$20^\circ$. While the solid curve represents the distribution derived
from the MDI data set, the dashed curve is a fit with a Gaussian fit
model with 4 free fit
parameters: position, amplitude, Gaussian width, and vertical
zero-point offset. This offset, which mimics the extended damping
wings of the distribution in the form of a flat, isotropic background,
becomes significant for the smallest bipolar regions, which have a
much wider angular distribution with a greatly elevated
background. 

While this choice of fit model may not give the best representation of
the {\it shape} of the extended wings of the distribution, it gives an 
extremely robust determination of the position and distribution width 
(full width at half maximum, or FWHM), which is the aim of the
fit. Other choices, like Voigt functions, give less stable inversions
in situations with non-optimum statistics, which we get when we 
further subdivide the material into separate flux bins. Note that the FWHM is 
a function of {\it both} the Gaussian width and the background 
offset.

\subsection{Dependence on flux and bipolar moment}\label{sec:flux}
Having determined Joy's law in the form of Eq.~(\ref{eq:globaltilt}),
we next want to explore whether the tilt amplitude $\gamma_0$ depends
on region size. There are several ways in which we can quantify the
region size. We here use two choices: (1) the total flux
$F_{\rm tot}=F_+ +F_-$, and (2) 
the ``bipolar moment'' $M$, defined as $M={1\over 2}F_{\rm tot}\,S$,
where $S$ is the separation (in Mm) between the centers of gravity of
the two polarities, as measured along the great circle that connects
them. It may be tempting to call $M$ the ``magnetic moment'' of the
bipolar region, but since this term is much used and defined very
differently in classical and quantum electrodynamics, we avoid 
confusion by using the term ``bipolar moment'' for $M$. 

\begin{figure}
\resizebox{\hsize}{!}{\includegraphics{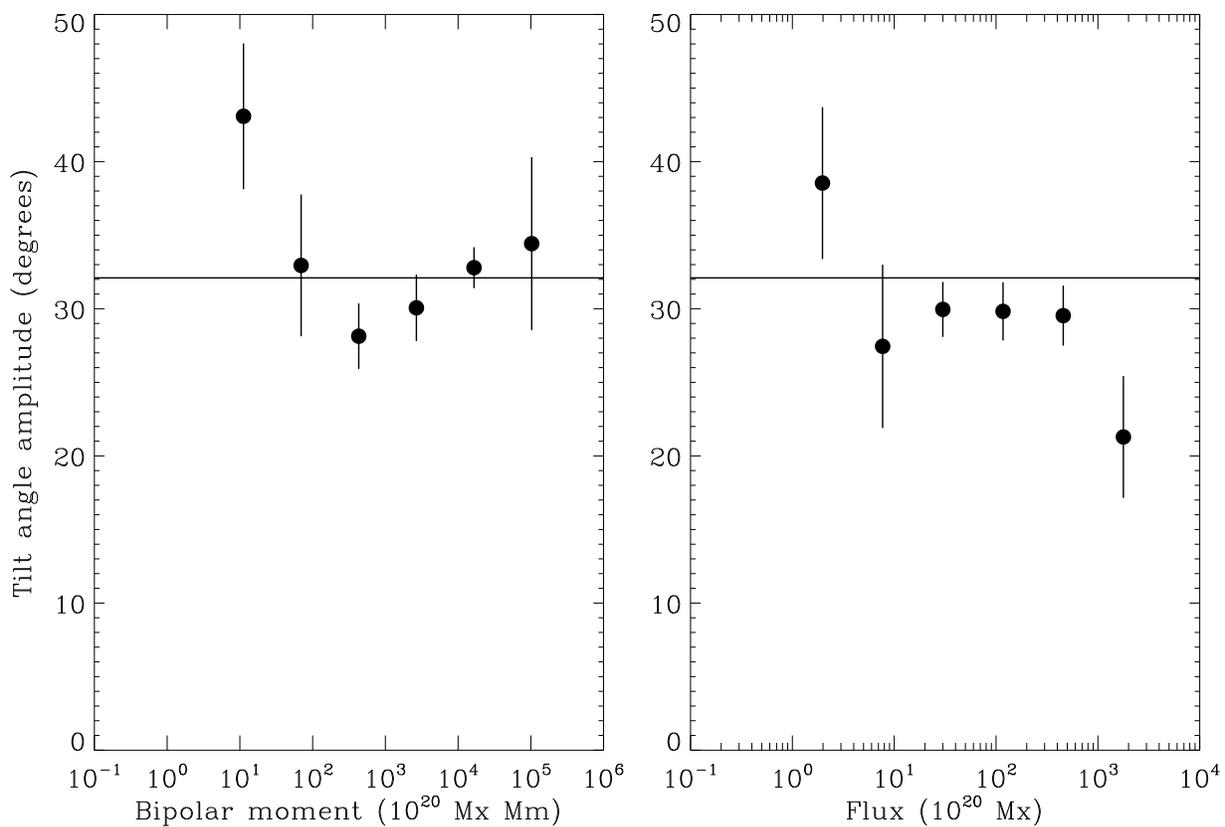}}
\caption{Tilt angle amplitude $\gamma_0$, defined by the fit function
  $\gamma=\gamma_0\sin b$, with the bipolar regions sorted in logarithmic
  bins with respect to their bipolar moments (left panel) and their
  total flux (right panel). The global
  average value of $\gamma_0$, as determined from Fig.~\ref{fig:globaltilt},
  is marked by the horizontal lines. Accounting for the error bars, we
  find no significant variation of tilt angle with flux or bipolar
  moment. 
}\label{fig:tiltbipmom}
\end{figure}

Since our data set spans several orders of magnitude in 
$F_{\rm tot}$ and $M$, we use 6 logarithmic bins for each of these two 
quantities. Figure \ref{fig:tiltbipmom} shows the result when the tilt
angle amplitude $\gamma_0$ is determined for each such 
bin. The $\gamma_0$ value of 32.1$^\circ$ from the global tilt analysis
of Fig.~\ref{fig:globaltilt} is drawn as a horizontal line for
reference. 

Figure \ref{fig:tiltbipmom} is consistent with the hypothesis that the
tilt angle amplitude is independent of flux and bipolar
moment. It might seem that the points for the smallest bipolar moment
bin and for the largest flux bin may indicate a negative slope (larger
average tilt for smaller regions), but as these points only deviate by
between 2 and 2.5 standard deviations from the global average, they
cannot be taken as evidence for a significant slope. 

Although we here cover several orders of magnitude in region size, we
should remember that our
lower flux limit is at about $10^{20}$\,Mx, just above the size range of
the ephemeral active regions. We have no evidence whether or not the
scale invariance of the average tilt also applies to these
smaller scales. 

We note that
according to the theory of \citet{stenflo-fanetal94}, which explains the
tilt in terms of the Coriolis force acting on flux loops that are
buoyantly rising from the bottom of the convection zone, the tilt
should vary with the total flux to the power of 1/4, which in
Fig.~\ref{fig:tiltbipmom} would correspond to a steep {\it positive} 
slope. With such a scaling law the tilt would increase by an order of
magnitude as we go from our smallest to our largest flux bin. Clearly
this theory can be ruled out by our results in Fig.~\ref{fig:tiltbipmom}. 

\begin{figure}
\resizebox{\hsize}{!}{\includegraphics{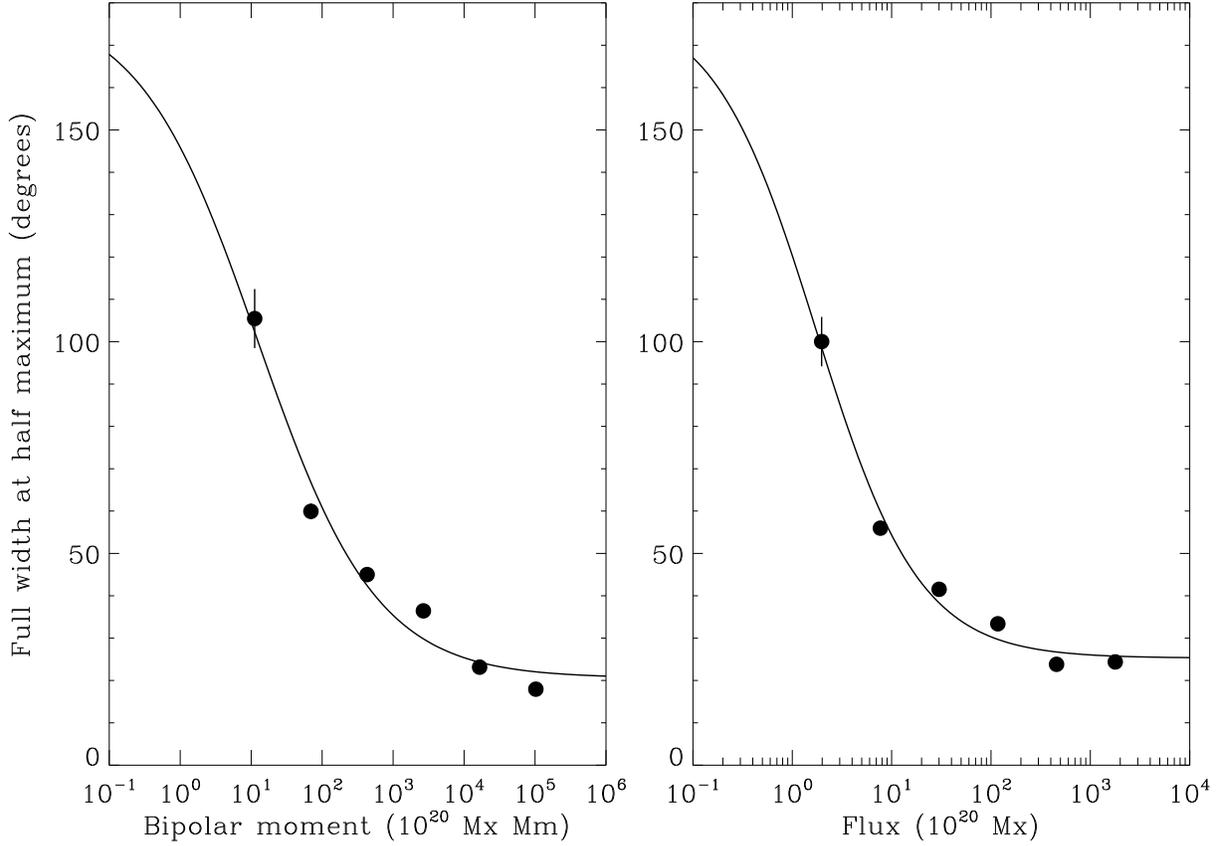}} 
\caption{Full width at half maximum of the tilt distributions for
  various bins in the logarithmic bipolar moment (left panel) and the
  logarithmic flux (right panel). The solid curves are fits with the
  analytical function $y=(180-a_2)a_0\,/(a_0+x^{a_1}\,)\,+a_2$, where
  $y$ is the width in degrees, and $x$ is either the bipolar moment or
  the flux. The values of the fit parameters are given in the text. 
}\label{fig:tiltspread}
\end{figure}

Figure \ref{fig:tiltspread} shows the full width at half maximum
(FWHM) of the angular tilt distributions as a function of logarithmic
bipolar moment (left panel) and logarithmic flux (right panel). The
plotted results represent averages over all latitudes. Thus the FWHM
with its error bar has been determined from the Gaussian fit model for
the angular distributions in each latitude bin (in
combination with the binning in $M$ and $F_{\rm tot}$). Then the
results for the different latitude bins have been averaged,
using the inverse of the squared error bars as weights. This gives us the
filled circles in the figure. 

The solid curves in the figure are analytical fit functions. We have chosen the
analytical form of these functions such that they smoothly
approach the value $180^\circ$ in the limit of zero $F_{\rm tot}$ or
$M$, because in this limit we may expect the distribution to become
nearly isotropic, while  in the asymptotic limit of large $F_{\rm tot}$ or 
$M$ a constant value for the FWHM may be reached (an assumption that
might not necessarily be valid). Such a fit
function needs 3 free parameters to give us sufficient freedom to find
good fits to the empirically
determined points. The choice that has been used for Fig.~\ref{fig:tiltspread}
has the form 
\begin{equation}
y=(180-a_2)a_0\,/(a_0+x^{a_1}\,)\,+a_2\,,\label{eq:fwhm}
\end{equation}
where $y$ represents the FWHM in degrees, and $x$ either stands for
$M$ or $F_{\rm tot}$. 

The fit curve in the left panel of Fig.~\ref{fig:tiltspread}, where
$x=M$ in units of $10^{20}$ Mx Mm, is defined by the following values
of the 3 free parameters: $a_0=3.7$, $a_1=0.52$, and
$a_2=20.6^\circ$. In the case of the right panel, where $x=F_{\rm tot}$ in 
units of $10^{20}$ Mx, the parameter values are: $a_0=1.59$, $a_1=0.84$, and
$a_2=25.3^\circ$. The somewhat different values (of $a_2$) in the
asymptotic limit of large $x$ is a reflection of the degree of
uncertainty with this choice of fit model with its implicit
assumptions. The model should be seen as a useful way to 
express the empirical behavior in compact form. 

Figure \ref{fig:tiltspread} shows that the largest bipolar regions
have a FWHM of their angular tilt distributions of $20^\circ$ -
$25^\circ$, but the angular spread increases dramatically as we go to
smaller regions, with an indication of becoming nearly isotropic
in the limit of zero region size. Note that the described behavior
represents an average of the FWHM determined separately for each
latitude. We have found no evidence for any significant dependence of the FWHM
on latitude.


\section{Violations of Hale's polarity law}
In the previous section with our exploration of Joy's law and the angular
distributions of the tilt angles we optimized the statistics by
shifting all tilt angles that fell outside
  the range [$-90^\circ,+90^\circ$] by $180^\circ$ to bring them back
  into this range, to allow bipolar regions from both
  hemispheres and different solar cycles to be combined into one
  single set of tilt
  distributions, which could then be studied as a function of latitude,
  flux, or bipolar moment. Through this optimization with $180^\circ$ shifts
  we erased information on Hale's law for the polarity orientations. 

In contrast, we will in the present section focus on Hale's polarity
law and let the tilt angles remain in any quadrant in which
they fall, without any $180^\circ$ shifts. Hale's polarity law states
that the polarity orientations of the bipolar magnetic regions is
opposite in the N and S hemispheres, and that the orientations in both
hemispheres get reversed when we pass from one 11-yr
cycle to the next.

\subsection{Butterfly diagrams for the bipolar orientations}
A common and effective way to visualize how the pattern of solar
activity evolves is in terms of the distribution of sunspots in
latitude-time space. This representation is called ``butterfly
diagram'', since the pattern reminds of butterflies that fly along the
time axis. Instead of sunspots we can let solar activity be
represented by bipolar magnetic regions. Having the tilt angles of the
bipolar regions we can go a step further and do butterfly-type
diagrams for bipolar magnetic regions with different
orientations. This allows us to illustrate Hale's polarity law in an
explicit graphical way. 

\begin{figure}
\resizebox{0.6\hsize}{!}{\includegraphics{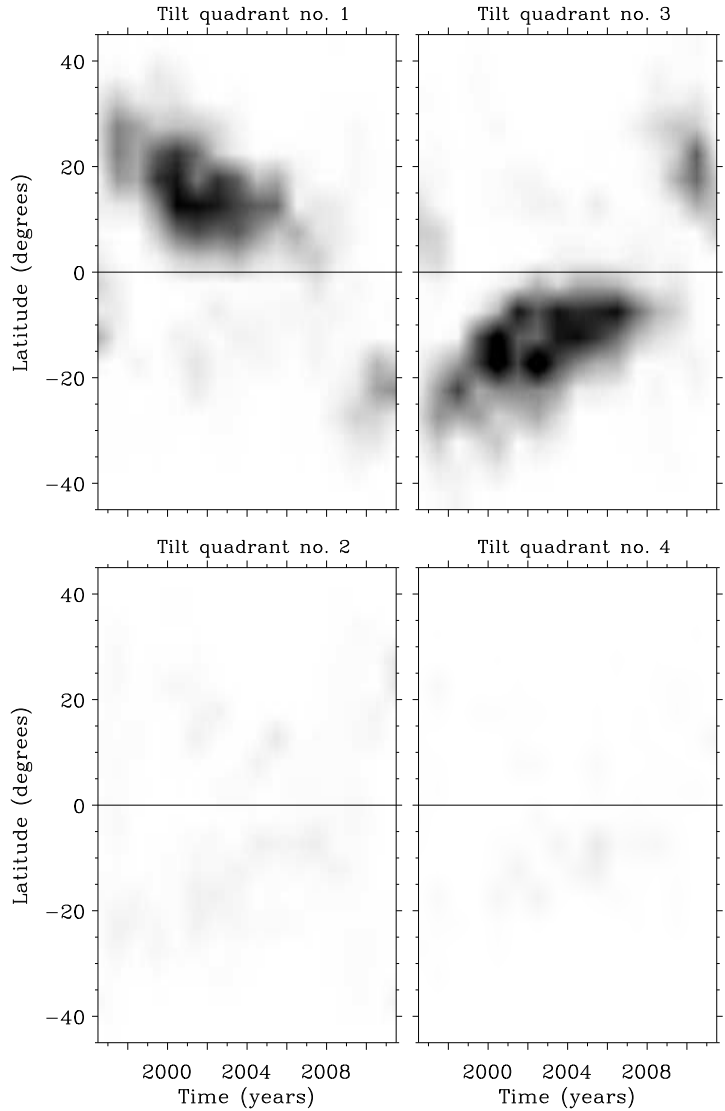}} 
\caption{Butterfly diagrams for the number of bipolar magnetic regions
  with tilt angles in each of the four quadrants. Quadrant no.~1 is
  centered around the average tilt angle for that latitude as given by
  the $\sin$(latitude) relation in Fig.~\ref{fig:globaltilt}. Quadrants 2, 3, and 4 are
  turned  $90^\circ$, $180^\circ$, and $270^\circ$ from this
  orientation. The grey-scale cut has been chosen such that black is at 80\,\%\ of
  the maximum value, white at zero. 
}\label{fig:butflynum}
\end{figure}

In Fig.~\ref{fig:butflynum} we have plotted the number density of
bipolar magnetic regions in latitude-time space separately for each of
the four tilt-angle quadrants, using all the regions in our data set,
regardless of their size. The quadrants are here defined with respect
to the nominal values of the tilt angles for the various latitudes
given by Joy's law, in the following way: If a given bipolar
region at latitude $b$ has a tilt angle that falls somewhere within
the $90^\circ$ interval $32.1^\circ \sin b \pm 45^\circ$,
then it is assigned to quadrant no.~1. It implies that its positive
polarity is the leading (westward) polarity. Quadrant no.~2 is
obtained by adding $90^\circ$ to the boundaries of quadrant no.~1,
while quadrants 3 and 4 are obtained by adding $180^\circ$ and
$270^\circ$, respectively. Thus bipolar regions belonging to quadrant
no.~3 have the same tilts as in quadrant no.~1, but with reversed
orientation, such that the negative polarity is the leading polarity. 

According to Hale's polarity law we expect quadrants 2 and 4 to
be unpopulated, while quadrants 1 and 3 should be populated in a
mutually exclusive way. Thus, for a given solar cycle, only one of the
N or S hemispheres should be populated in quadrant no.~1, while only
the opposite hemisphere should be populated in quadrant no.~3. As the
next cycle begins, the hemispheres that are populated get exchanged. 

The pattern shown in Fig.~\ref{fig:butflynum} illustrates these
properties of Hale's polarity law in a very explicit way, while
indicating that it is a law that is not strictly obeyed but is fairly 
often violated. The familiar active-region belts that migrate from
higher to lower latitudes are clearly seen, but only
in the N hemisphere in quadrant no.~1 (before 2008), while the S
hemisphere belt occurs in quadrant no.~3. After 2008 this pattern
reverses as the new cycle starts. In comparison, quadrants 2 and 4 are
nearly empty, as expected from Hale's law. 

The grey-scale cuts in Fig.~\ref{fig:butflynum} 
have been chosen such 
that white represents zero, black 80\,\%\ of the
global maximum value. While Hale's law is obeyed in the
great majority of cases, we notice that quadrants 2 and 4 are not
entirely empty, and that the ``wrong'' hemispheres in quadrants 1 and 3 are
also weakly populated. As these apparent violations of Hale's law
appear to be rather randomly distributed, the question arises to what extent
they represent noise fluctuations and errors in our automatic region
identifications, or are real, physical violations of Hale's law. We
address this question in the next section.

\subsection{Examples of unambiguous violations}
\begin{figure}
\resizebox{\hsize}{!}{\includegraphics{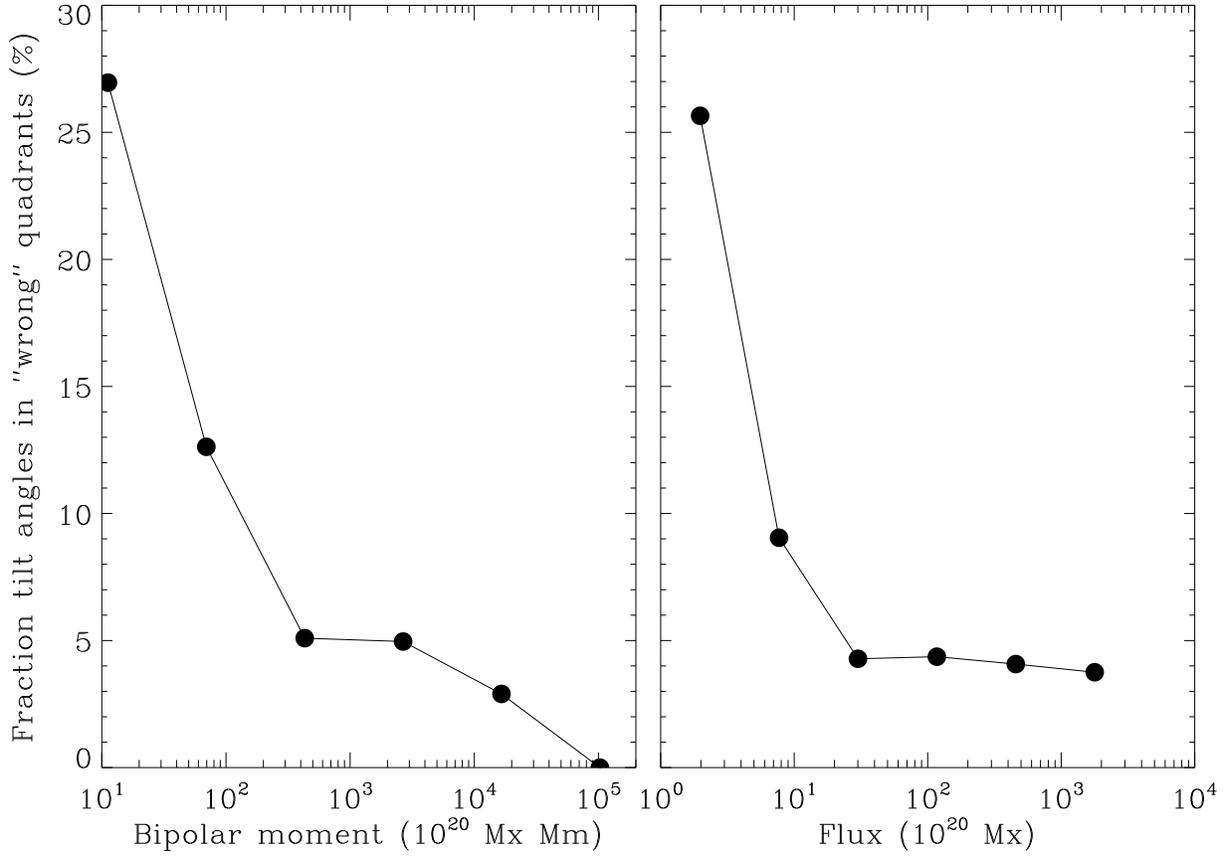}}
\caption{Fraction of bipolar magnetic regions with tilt angles in the
  ``wrong'' quadrants (i.e., quadrants 2 and 4 in
  Fig.~\ref{fig:butflynum}). Left panel: as function 
  of logarithmic bipolar moment bin. Right panel: as function of logarithmic
  flux bin. 
}\label{fig:wrongquad}
\end{figure}

Let us first note that the frequency of violations of Hale's polarity
law increases rather steeply as we go to regions of smaller size. This
behavior is illustrated in  Fig.~\ref{fig:wrongquad}, where we have
plotted the fraction of bipolar regions assigned to the ``wrong''
quadrants 2 and 4, relative to the total number of regions, as
functions of bipolar moment (left panel) and flux (right panel). For
medium-size and large bipolar regions this fraction is typically
4\,\%, and possibly less for the largest regions (although here
the statistics is poor). This agrees with
the conclusions of \citet{stenflo-richardson48},
\citet{stenflo-wangsheeley89}, \citet{stenflo-khlystova2009}, and
\citet{stenflo-sokoloff2010}, who find similar frequencies of
violations of Hale's law. For our smallest
size bin, however, the violating fraction exceeds 25\,\%\ with a very steep
gradient, indicating that the tilt distribution may get randomized in
the limit of zero region size. 

This behavior is consistent with our findings for the FWHM of the tilt
distributions in Fig.~\ref{fig:tiltspread}, which showed a dramatic
increase of the distribution width as we go to smaller regions. As the
angular distributions of the smaller regions not only 
have larger half widths but also extended wings that appear like an
elevated, nearly isotropic background, there is no wonder that there
will be significant spill-over into the ``wrong'' tilt quadrants. 

Figure \ref{fig:wrongquad} does not distinguish between quadrants 1
and 3, because to do this one would need to assign the right cycle
number to a given region, which cannot be done in a fully unambiguous
way. However, an approximate treatment indicates that the fraction of
regions that fall in quadrant 1 when it should belong to quadrant 3,
and vice versa, is similar to the fractions that fall in the other
wrong quadrants 2 and 4. 

The circumstance that the tilt distributions have extended wings and
elevated backgrounds with significant spill-over into the wrong
quadrants could to some degree be due to errors from 
misidentifications, in particular in crowded magnetograms. Visual
inspection of magnetograms for a sample of violating cases however
indicates that the majority of the violations represent a real
property of the Sun. Since in the early phase of active-region
development, during the first days after flux emergence, the tilt
orientation is often found to rotate until it settles to a value more
in agreement with Joy's law \citep{stenflo-ks08}, and since
we in the present analysis do not include information on the age of a
bipolar region, it could be that evolutionary rotation of bipolar
regions may contribute to the spread of tilt angles. 

\begin{figure}
\resizebox{\hsize}{!}{\includegraphics{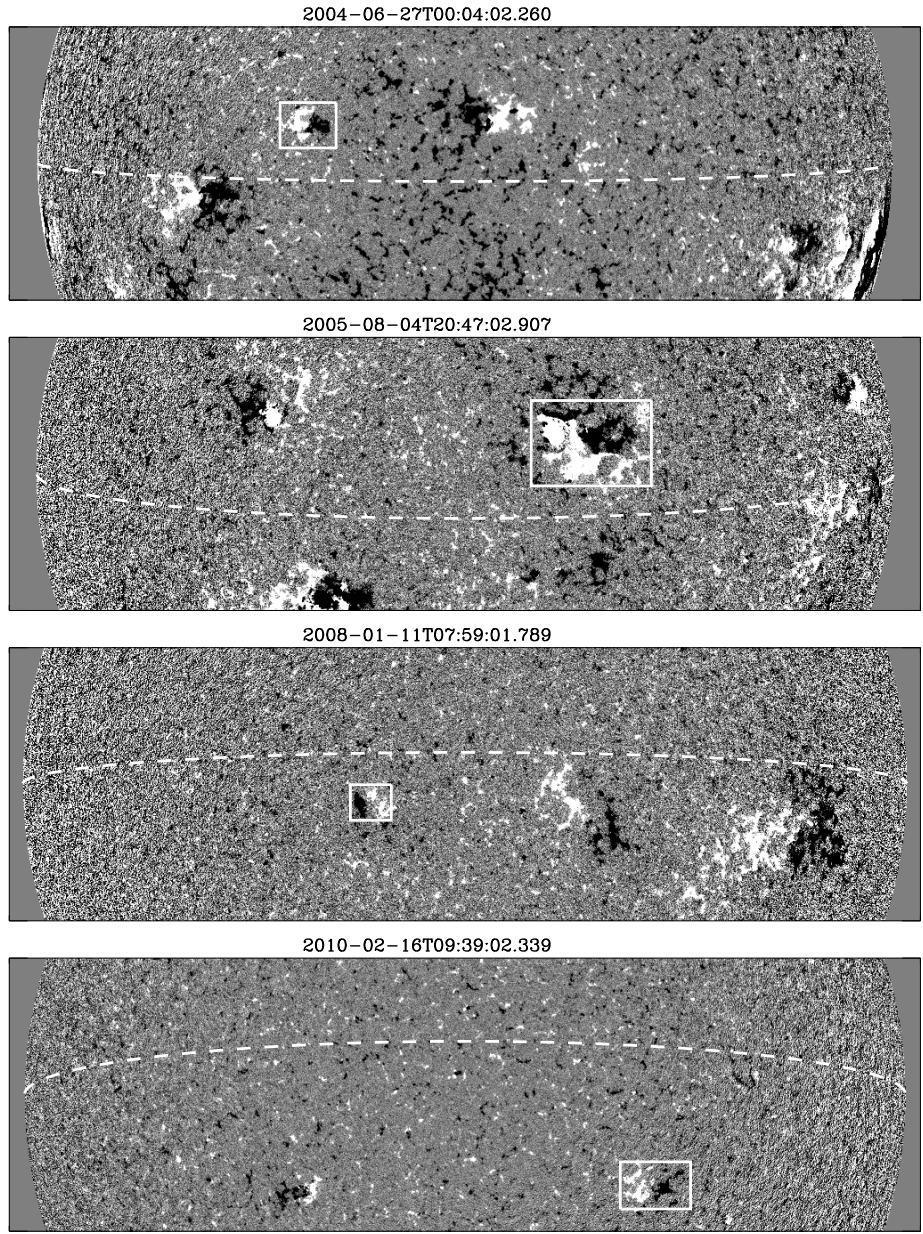}}
\caption{Four explicit examples, marked by the enclosing rectangular
  boxes, where Hale's polarity law is being violated in an unambiguous
  way by medium-size or large bipolar magnetic regions. In all four
  cases there are well-defined bipolar regions 
  that obey Hale's law in the same latitude zone of the same
  magnetogram. The location of the solar equator is marked by the
  dashed line. The chosen examples are spread over various phases of
  the solar cycle: 27 June 2004, 4 August 2005, 11 January 2008, 16
  February 2010. The grey-scale cuts are set at $-50$ (dark) and
  $+50$\,G (white). 
}\label{fig:violate}
\end{figure}

To explicitly demonstrate in an unambiguous way that many physical,
undisputable violations of Hale's polarity law do indeed exist, not only
for small bipolar regions but also for large ones, we show in Fig.~\ref{fig:violate} four
cases (enclosed by rectangular boxes), selected from different
phases of the solar cycle (27 June 
2004, 4 August 2005, 11 January 2008, 16 February 2010), where a large
bipolar region has reversed orientation (being wrong by approximately $180^\circ$),
while in the same magnetogram a prominent bipolar region with the
correct polarity orientation is present in the {\it same latitude}
  strip. The heliographic latitudes of the four violating regions are $+7.6^\circ$, $+10.3^\circ$,
$-6.7^\circ$, and $-18.4^\circ$.  We have chosen these examples such 
that the violating and the 
  non-violating regions should be at least medium-size and side by
  side in the same latitude 
  strip, while at the same time being well separated from each other and all
  other surrounding regions, so that there can be no question about their
  identifications as distinct, individual bipolar magnetic
  regions. Inspections of time sequences of magnetograms confirm that
  the violations are not due to rotations during region evolution, but
  are really stable properties of the regions. 

The violating case illustrated for 16 February 2010 has previously
been reported in \citet{stenflo-spw6s11}. Due to the latitude-time positions and
magnitudes of these violating regions, we can exclude the possibility
that they belong to the ``wrong'' solar cycle. These violations 
are clearly incompatible with  
a Babcock-Leighton type scenario
\citep{stenflo-babcock61,stenflo-leighton69}, according to which one
subsurface toroidal flux 
belt is the source of the emerging bipolar magnetic regions. If major, 
stable bipolar regions with opposite polarity orientations appear at the same
time in the same latitude zone, they cannot possibly be part of the
same toroidal flux system. 

The reported violations of Hale's polarity law may be seen as
anomalies, but the discovery of such anomalies helps expose deep
problems within a given paradigm.


\section{Conclusions}\label{sec:concl}
The bipolar magnetic regions represent the most conspicuous directly
observable signatures of the dynamo that operates in the Sun's
interior. While their general E-W polarity orientations indicate that
they represent amplified largely toroidal flux, their systematic
tilt shows that the emerging flux also brings to the surface an N-S dipole
moment that represents the seed of the new poloidal field that will
replace the old one, with reversed orientation. The tilt angles thus
represent explicit signatures of the dynamo 
process that is responsible for the regeneration of the poloidal field
from the toroidal one. 

A prevailing paradigm for the origin of the tilt angles has been in
terms of the Coriolis force acting on flux loops that buoyantly rise
from the tachocline region at the bottom of the convection zone, where the
dynamo is assumed to operate 
\citep{stenflo-dsilva93,stenflo-fanetal94,stenflo-fisheretal95}. This paradigm requires
superstrong (60-100\,kG) toroidal magnetic fields without appreciable
tilt in the tachocline zone, much stronger than the value for
equipartition with the convective motions. As the speed of the buoyant
rise depends both on the amount of flux and the strength of the field,
one would expect that the 
observed tilt at the surface should depend on the size of the bipolar regions. 

Work by
\citet{stenflo-sivaramanetal07} and \citet{stenflo-ks08} has however
shown that this paradigm is untenable, since the tilt angles are found
to relax after emergence not towards the E-W orientation but towards the angle
prescribed by Joy's law, and this behavior is found to be independent
of size or amount of flux of the regions. This is evidence that the
tilt (or N-S dipole moment) is already established in the source region
inside the Sun and not during the buoyant rise of the flux loops. The
present work supports this conclusion by showing, with much enhanced
statistics, that the average tilt does not change appreciably as we go to
regions with orders of magnitude smaller flux contents. 

An even much more ingrained, long-term paradigm that is shown to be
untenable by the present work is the phenomenological scenario, according to
which differential rotation creates from a poloidal field a coherent
toroidal flux system, from which the sunspots arise. Such a scenario
is not compatible with the violations of Hale's polarity law that we
have presented here. Our illustrated examples where well defined
medium-size bipolar magnetic regions occur side by side {\it in the
  same magnetogram and the same latitude zone} unambiguously show
that these oppositely oriented regions cannot be part of the same flux
system, but that there must be a coexistence of oppositely oriented
toroidal flux in the same latitude zones at any given time. 

Although the occurrencies of violations of Hale's polarity law represent only
a few percent of all cases, they rule out the
possibility of well defined, coherent toroidal flux systems as a
source of all active regions, even the large ones. Our results make
clear that {\it fluctuations} represent an essential inherent physical
property of the solar dynamo, as expected to various degrees from
turbulent dynamo theory \citep{stenflo-brandsub05}. 

\begin{figure}
\resizebox{\hsize}{!}{\includegraphics{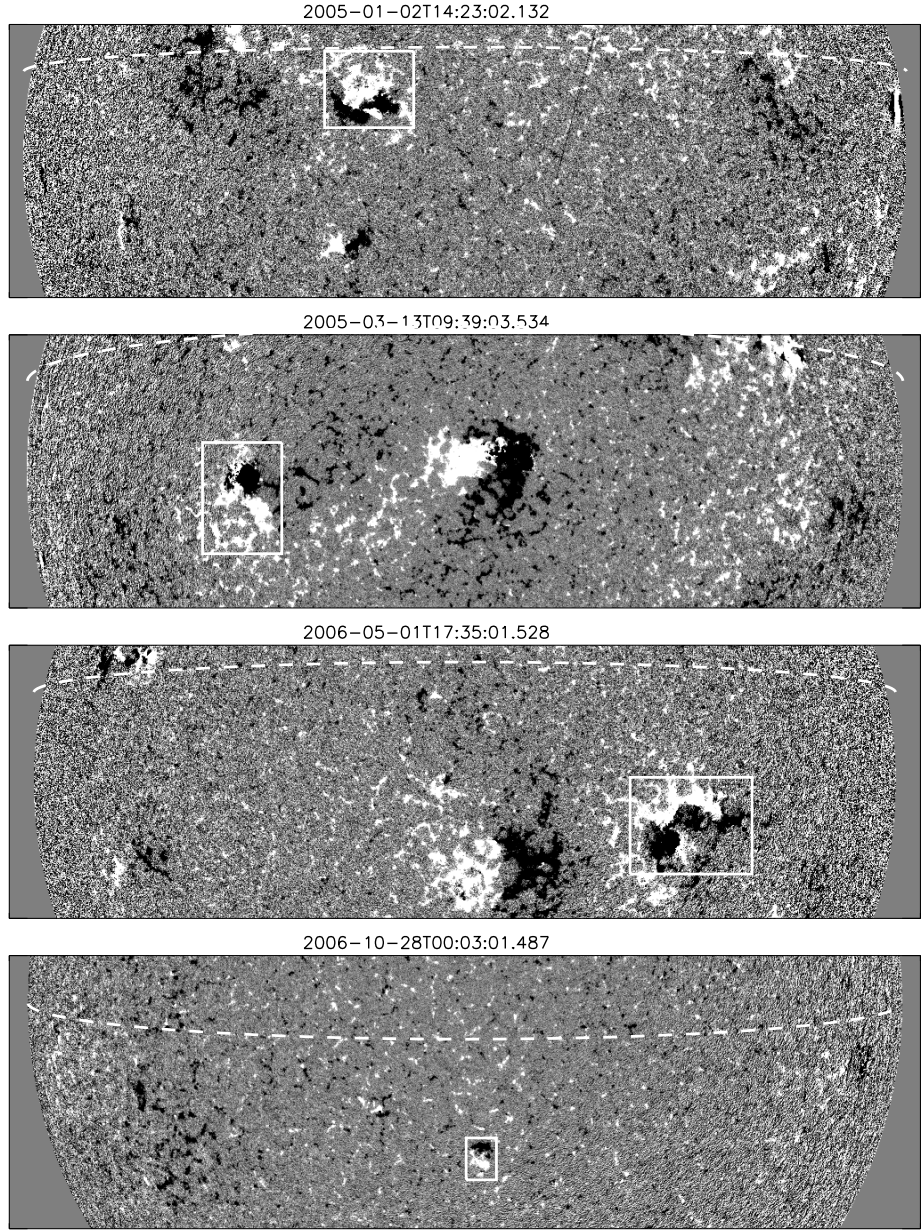}}
\caption{In contrast to Fig.~\ref{fig:violate}, which showed examples
  of regions with orientations differing by nearly $180^\circ$ from
  the orientations expected from Hale's and Joy's laws, we here show
  four examples where the orientation differs by an amount between
  $90^\circ$ and $100^\circ$ (marked by the enclosing rectangular
  boxes). The location of the solar equator is marked by the
  dashed line. The grey-scale cuts are set at $-50$ (dark) and
  $+50$\,G (white). 
}\label{fig:90deg}
\end{figure}

The fundamental role of large fluctuations at all scales is further
illustrated in Fig.~\ref{fig:90deg} for four selected cases of well
defined bipolar magnetic regions, including large ones, which have 
orientations that differ by $90^\circ$ - $100^\circ$ from the
orientations prescribed by Joy's law. The existence of many such cases
shows that we have fluctuations over the whole range of orientation
angles, not only between ``Hale and anti-Hale'' (proper and reversed)
orientations. Also the largest bipolar regions are subject to these
fluctuations. 

Our results showing the profound role of fluctuations together with the
finding that the average tilt angles described by Joy's law are
independent of the amount of flux of the regions, over several orders
of magntiude in flux, cast doubt on the validity of tachocline dynamo
theories, and seem to be more compatible with distributed dynamo
models \citep[cf.][]{stenflo-brandenburg05}. 

Other concepts that are not supported by our analysis is the
separation of scales and the separation between a global and a local
dynamo. Previous work has shown that the size spectrum of magnetic
regions follows a power law indicating scale invariance over the whole
range of resolved scales \citep[cf.][]{stenflo-parnelletal09}. Our
finding that the average tilt angle is independent of region size
shows that also the smallest regions in our sample contribute
significantly to the N-S dipole moment that leads to the reversal of
the global dipole field. There is no observational evidence that below
a certain scale size the fields no more contribute to the global
dynamo. On the contrary the accumulated global effect of the small
regions may even be the dominating one, because it has long been well
known that the smallest regions dominate the global flux emergence
rate \citep{stenflo-zirin87}. Since our analysis indicates that even
the smallest regions, although increasingly randomized in their
orientations, retain a non-random component, their net global effect
may be substantial. Although it is beyond the scope of the present
paper to quantify the relative global contributions of the various
scales, we feel that the artificial separation of scales and division
between a global and local dynamo are theoretical idealizations that
are not justified by observations of the Sun.


\acknowledgments
We wish to acknowledge the fruitful discussions about some of these issues that
took place at ISSI (International Space Science Institute) in Bern
during an international team meeting, November 15-19,
2010, and during a Nordita Workshop on ``Dynamo, Dynamical Systems
and Topology'' in Stockholm in August 2011. One of us (J.O.S.) wants
to acknowledge the hospitality of Stanford University during a visit
in March 2011 for work on the data set recorded by the MDI instrument on the SOHO
spacecraft. SOHO is a project of 
international cooperation between ESA and NASA. 


\appendix
\section{Reduction steps}\label{sec:steps}
The identification algorithm has been developed in a pragmatic way,
with the selection criteria and thresholds represented by a set of
free parameters. The values of these free parameters have been manually
optimized and then fixed, by comparing the program identifications
with visual inspections for a selection of magnetograms representing
all phases of solar activity, from the most active and crowded phase,
to the most quiet phase. The parameters have been optimized such that the
same parameter set can be used for all the magnetograms, for all phases of
the solar activity cycle. 

Once this manual program optimization has been done, the rest is
automatic. The optimized IDL program is run in batch mode in a loop
analysing each of the 73,838 MDI magnetograms one by one. For each
analysed magnetogram an IDL save file is written that contains all 
the extracted bipolar region parameters with the relevant housekeeping
data. The effective computing time needed to run this large IDL batch job at
Stanford was about 2 weeks. 

\subsection{Step 1}\label{sec:step1}
Each magnetogram is analysed in a sequence of program steps, done by
separate subroutines. In Step 1 the line-of-sight magnetogram is
converted to a magnetogram for the vertical flux density $B_v$, simply by
dividing the line-of-sight component $B_\parallel$ with $\mu$, the
cosine of the heliocentric angle, except in the limb zone with radius
vector $r/r_\odot$ between 0.9 and 1.0, in which we divide by 1/2.29,
the $\mu$ value for $r/r_\odot=0.9$, to avoid excessive amplification
(of errors and noise) near the limb. Since we later only retain
bipolar regions within $r/r_\odot<0.8$, the way in which we treat the
limb zone is rather irrelevant for the later bipolar region analysis. 

The number of defect pixels (with NaN values) are counted in
Step 1 and replaced by zeros. If the number of such pixels on the
solar disk exceeds 100 (representing about 0.01\,\%\ of the total
number), the magnetogram is rejected. Furthermore the average of the
unsigned vertical flux density $\vert B_v\vert$ within $r/r_\odot<0.9$
is determined (and here called $B_{\rm ave}$), since it is needed in
Step 3. 

In Step 1 we also produce a spatially smoothed version of $B_v$,
and in this smoothed version set everything that has an absolute value
below a certain threshold $B_{\rm cut}$ to zero (thus only retaining the ``tips of
the icebergs'' in this smoothed magnetogram). For later reference we
call this smoothed and cut magnetogram $B_{\rm sm}$. The bipolar region identification is
based on this version, to filter out the large number of 
magnetic elements that have high flux density values only over a small
number of neighboring pixels, since inclusion of them would lead to
confusion in the overall identification also for the larger
regions. The main reason for the smoothing is that the larger regions
usually have a substantial spatial gap of low 
flux density between the two opposite polarities, and this gap gets
reduced and is more easily bridged in the smoothed and cut 
magnetograms. 

Through trial and error we have determined the optimum choice for the 
width of the square smoothing window to be 11 pixels or
22\,arcsec. For the choice of cut $B_{\rm cut}$, below which the
smoothed field is set to zero, we have found it necessary to link it
to the overall flux level $B_{\rm ave}$ to 
enable automatic bipolar region identification with a single algorithm
that works for all phases of solar activity. This link is done by the setting  
$B_{\rm_cut}=5.8(B_{\rm ave}-12)$ or 30\,G, whichever is larger. As we will see under Step 3, this
initial choice for $B_{\rm cut}$ is adjusted incrementally upwards to avoid 
faulty identifications in crowded fields, if certain criteria are not satisfied. 

While these choices allow good identifications of the 
large bipolar regions, they obviously lead to a filtered suppression of the
contributions from small regions. However, since this suppression does
not lead to a sharp cut-off, while the number of bipolar regions
increases steeply with decreasing size, our 
smoothing procedure still gives us a statistically very useful sample of small-scale bipolar
regions, although the sample is incomplete. 

\subsection{Step 2}
In Step 2 we use the unsigned version $\vert B_{\rm sm}\vert$ of the
smoothed and cut magnetogram, identify each ``island'' (contiguous
region of non-zero pixels), and draw a rectangular box around each
such ``island''  with a margin of fixed width on all sides. Through
trial and error optimization the width of this margin has been chosen
to be 9 pixels or 18\,arcsec, meaning that the rectangular boundary box is at
least 18\,arcsec beyond the island pixels on all sides. We need a
certain margin width to bridge the gap that frequently occurs between
the two polarities in a bipolar region, but if the margin is chosen 
too wide, we get excessive overlap between the different region boxes
(which adversely affects the consolidation process described under
Step 3). 

\subsection{Step 3}
In the consolidation process of Step 3 we go through all the
``islands'' identified under Step 2 and check for overlaps between
their respective region boxes. When two region boxes partially
overlap, they are replaced by a single rectangular box that encloses
both of them. This process of mergers is continued until there is no
more overlap between boxes. The resulting number of regions is thereby
greatly reduced. 

During times of high levels of solar activity, when the magnetogram is
crowded with neighbouring prominent bipolar regions, our
process of merging partially overlapping region boxes may produce 
boxes that are too large, because they enclose not only a single
bipolar region but also some of its neighbours. This can lead to faulty
results, since the extraction of bipolar region properties under Step
4 is done under the assumption that each region box contains a
single bipolar region. Through trial and error we have found the following way
to deal with such cases to minimize the number of faulty identifications: 

If a merged region box has a width exceeding 240 pixels in the E-W direction,
then there is a significant chance that it does not represent a single
bipolar region. In such a case the value of the cut parameter
$B_{\rm_cut}$ is raised by 10\,\%\ to produce a new version of the
smoothed 
and cut magnetogram $B_{\rm sm}$, and one goes back to Step 2 with
this new $B_{\rm sm}$ as input. If with this modification a merged
region box still exceeds a width of 240 pixel, then the same procedure
with raising the value of $B_{\rm_cut}$ by 10\,\%\ and returning to
Step 2 is repeated. This repetitive incremental increase of
$B_{\rm_cut}$ can continue for a maximum of 7 times, until the box
width criterion is satisfied. If the criterion remains unsatisfied,
the magnetogram is discarded. 

When the box width criterion is satisfied, we loop through the
remaining, merged region boxes in the magnetogram and apply two
additional criteria to decide whether or not the contents of a given region box
can qualify as a bipolar magnetic region. The first
criterion is that both the positive and negative amplitudes of the unsmoothed vertical flux
density $B_v$ should be large enough. Thus we require the maximum value
of $B_v$ to be larger than 200\,G and the
minimum value smaller than $-200$\,G. 

The second criterion is that the fluxes
of positive and negative polarities within a box should be
moderately balanced, since a box with too much dominance of a single
polarity cannot be classified as bipolar. Let flux $F_+$ be the sum of
all the $B_v$ values for pixels with $B_v > 100$\,G, and similarly $F_-$ be the sum of
all the $B_v$ values for pixels with $B_v <-100$\,G. Our 
balance criterion is satisfied if $(F_+ +F_-)/(F_+ -F_-)\,<0.4$. For
perfect balance this quantity would be zero, for a monopolar region it 
would be 1.0. This criterion cannot be made too strict, since we know
that a large fraction of all truly bipolar regions have moderately
unbalanced polarities, with field-line links to other parts of the
Sun. 

\subsection{Step 4}
In Step 4 we loop through all the final region boxes that have passed
all our criteria to qualify as bipolar regions and derive their
properties. For each of the two polarities we derive the amount of flux
and the heliographic coordinates of their centers of gravity, using only the pixels within 
a box that have $\vert B_v\vert$ (unsmoothed vertical flux density) larger than 
100\,G. If $r/r_\odot$ for the centers of gravity is not $<0.8$ for
both polarities, then the bipolar region is not retained for analysis,
to avoid errors that can become magnified when analysing regions too
close to the limb. 

Using spherical trigonometry we draw a great circle through the
centers of gravity of the positive and negative polarities to
calculate the distance between them and their tilt angle with respect
to the E-W direction, as described in Sect.~\ref{sec:extract}.



\begin{thebibliography}{29}
\expandafter\ifx\csname natexlab\endcsname\relax\def\natexlab#1{#1}\fi

\bibitem[{{Babcock}(1961)}]{stenflo-babcock61}
{Babcock}, H.~W. 1961, \apj, 133, 572

\bibitem[{{Brandenburg}(2005)}]{stenflo-brandenburg05}
{Brandenburg}, A. 2005, \apj, 625, 539

\bibitem[{{Brandenburg} \& {Subramanian}(2005)}]{stenflo-brandsub05}
{Brandenburg}, A. \& {Subramanian}, K. 2005, \physrep, 417, 1

\bibitem[{{Dasi-Espuig} {et~al.}(2010){Dasi-Espuig}, {Solanki}, {Krivova},
  {Cameron}, \& {Pe{\~n}uela}}]{stenflo-espuig11}
{Dasi-Espuig}, M., {Solanki}, S.~K., {Krivova}, N.~A., {Cameron}, R., \&
  {Pe{\~n}uela}, T. 2010, \aap, 518, A7

\bibitem[{{de Wijn} {et~al.}(2009){de Wijn}, {Stenflo}, {Solanki}, \&
  {Tsuneta}}]{stenflo-dewijn09}
{de Wijn}, A.~G., {Stenflo}, J.~O., {Solanki}, S.~K., \& {Tsuneta}, S. 2009,
  Space Science Reviews, 144, 275

\bibitem[{{D'Silva} \& {Choudhuri}(1993)}]{stenflo-dsilva93}
{D'Silva}, S. \& {Choudhuri}, A.~R. 1993, \aap, 272, 621

\bibitem[{{Fan} {et~al.}(1994){Fan}, {Fisher}, \&
  {McClymont}}]{stenflo-fanetal94}
{Fan}, Y., {Fisher}, G.~H., \& {McClymont}, A.~N. 1994, \apj, 436, 907

\bibitem[{{Fisher} {et~al.}(1995){Fisher}, {Fan}, \&
  {Howard}}]{stenflo-fisheretal95}
{Fisher}, G.~H., {Fan}, Y., \& {Howard}, R.~F. 1995, \apj, 438, 463

\bibitem[{{Hale} {et~al.}(1919){Hale}, {Ellerman}, {Nicholson}, \&
  {Joy}}]{stenflo-haleetal19}
{Hale}, G.~E., {Ellerman}, F., {Nicholson}, S.~B., \& {Joy}, A.~H. 1919, \apj,
  49, 153

\bibitem[{{Harvey}(1993)}]{stenflo-harveyphd93}
{Harvey}, K.~L. 1993, PhD thesis, , Univ.~Utrecht, (1993)

\bibitem[{{Harvey} {et~al.}(1975){Harvey}, {Harvey}, \&
  {Martin}}]{stenflo-harveyetal75}
{Harvey}, K.~L., {Harvey}, J.~W., \& {Martin}, S.~F. 1975, \solphys, 40, 87

\bibitem[{{Harvey} \& {Martin}(1973)}]{stenflo-harveymartin73}
{Harvey}, K.~L. \& {Martin}, S.~F. 1973, \solphys, 32, 389

\bibitem[{{Harvey} \& {Zwaan}(1993)}]{stenflo-harveyzwaan93}
{Harvey}, K.~L. \& {Zwaan}, C. 1993, \solphys, 148, 85

\bibitem[{{Howard}(1991{\natexlab{a}})}]{stenflo-howard91b}
{Howard}, R.~F. 1991{\natexlab{a}}, \solphys, 136, 251

\bibitem[{{Howard}(1991{\natexlab{b}})}]{stenflo-howard91a}
{Howard}, R.~F. 1991{\natexlab{b}}, \solphys, 132, 49

\bibitem[{{Khlystova} \& {Sokoloff}(2009)}]{stenflo-khlystova2009}
{Khlystova}, A.~I. \& {Sokoloff}, D.~D. 2009, Astronomy Reports, 53, 281

\bibitem[{{Kosovichev} \& {Stenflo}(2008)}]{stenflo-ks08}
{Kosovichev}, A.~G. \& {Stenflo}, J.~O. 2008, \apjl, 688, L115

\bibitem[{{Leighton}(1969)}]{stenflo-leighton69}
{Leighton}, R.~B. 1969, \apj, 156, 1

\bibitem[{{Martin} \& {Harvey}(1979)}]{stenflo-martinharvey79}
{Martin}, S.~F. \& {Harvey}, K.~L. 1979, \solphys, 64, 93

\bibitem[{{Parnell} {et~al.}(2009){Parnell}, {DeForest}, {Hagenaar},
  {Johnston}, {Lamb}, \& {Welsch}}]{stenflo-parnelletal09}
{Parnell}, C.~E., {DeForest}, C.~E., {Hagenaar}, H.~J., {et~al.} 2009, \apj,
  698, 75

\bibitem[{{Richardson}(1948)}]{stenflo-richardson48}
{Richardson}, R.~S. 1948, \apj, 107, 78

\bibitem[{{Scherrer} {et~al.}(1995){Scherrer}, {Bogart}, {Bush}, {Hoeksema},
  {Kosovichev}, {Schou}, {Rosenberg}, {Springer}, {Tarbell}, {Title},
  {Wolfson}, {Zayer}, \& {MDI Engineering Team}}]{stenflo-scherreretal95}
{Scherrer}, P.~H., {Bogart}, R.~S., {Bush}, R.~I., {et~al.} 1995, \solphys,
  162, 129

\bibitem[{{Schrijver} \& {Harvey}(1994)}]{stenflo-schrharv94}
{Schrijver}, C.~J. \& {Harvey}, K.~L. 1994, \solphys, 150, 1

\bibitem[{{Sch{\"u}ssler} \& {Baumann}(2006)}]{stenflo-schbau06}
{Sch{\"u}ssler}, M. \& {Baumann}, I. 2006, \aap, 459, 945

\bibitem[{{Sivaraman} {et~al.}(2007){Sivaraman}, {Gokhale}, {Sivaraman},
  {Gupta}, \& {Howard}}]{stenflo-sivaramanetal07}
{Sivaraman}, K.~R., {Gokhale}, M.~H., {Sivaraman}, H., {Gupta}, S.~S., \&
  {Howard}, R.~F. 2007, \apj, 657, 592

\bibitem[{{Sokoloff} \& {Khlystova}(2010)}]{stenflo-sokoloff2010}
{Sokoloff}, D. \& {Khlystova}, A.~I. 2010, Astronomische Nachrichten, 331, 82

\bibitem[{{Stenflo}(2011)}]{stenflo-spw6s11}
{Stenflo}, J.~O. 2011, in Astronomical Society of the Pacific Conference
  Series, Vol. 437, Astronomical Society of the Pacific Conference Series, ed.
  {J.~R.~Kuhn, D.~M.~Harrington, H.~Lin, S.~V.~Berdyugina, J.~Trujillo-Bueno,
  S.~L.~Keil, \& T.~Rimmele}, 3--17

\bibitem[{{Wang} \& {Sheeley}(1989)}]{stenflo-wangsheeley89}
{Wang}, Y.-M. \& {Sheeley}, Jr., N.~R. 1989, \solphys, 124, 81

\bibitem[{{Zirin}(1987)}]{stenflo-zirin87}
{Zirin}, H. 1987, \solphys, 110, 101

\end{thebibliography}
\end{document}